Enhancement of hole mobility in high-rate reactively sputtered $Cu_2O$ thin films induced by laser thermal annealing

Jiří Rezek[1]*, Martin Kučera[2], Tomáš Kozák[1], Radomír Čerstvý[1], Aleš Franc[2], Pavel Baroch[1]

[1]Department of Physics and NTIS, European Centre of Excellence, University of West Bohemia, Univerzitní 8, 301 00 Plzeň, Czech Republic

[2]New Technologies – Research Centre, University of West Bohemia, Univerzitní 8, 301 00 Plzeň, Czech Republic

*Corresponding author, Tel.: +420 377632269, E-mail address: jrezek@ntis.zcu.cz

Abstract

In presented work, a reactive high-power impulse magnetron sputtering (r-HiPIMS) was used for high-rate deposition ($\approx$ 170 nm/min) of $Cu_2O$ films. Films were deposited on a standard soda-lime glass (SLG) substrate at a temperature of 190 °C. As-deposited films exhibit poor hole mobility in the orders of $\approx$ 1 $cm^2$/Vs. We have systematically studied the effect of laser thermal annealing (LTA) procedure performed using high-power infrared laser under different laser parameters (number of pulses, length of the pulse). We have found, LTA procedure could significantly enhance the hole mobility (up to 24 $cm^2$/Vs in our case). We have also fitted the results of a temperature-dependent Hall measurement to clarify the mechanism of the reported increase in hole mobility. Moreover, we have discussed the effect of the LTA procedure on microstructure (crystallinity, surface morphology) and on the value of optical band gap.

Keywords: $Cu_2O$; reactive HiPIMS; laser thermal annealing; hole mobility

1. Introduction

The cuprous oxide-based ($Cu_2O$) materials are one of the most promising and studied p-type transparent conductive oxide (TCO) materials. It is because of copper earth-abundancy, $Cu_2O$ non-toxicity and relatively suitable optoelectrical properties. However, the performance of all state-of-the-art p-type TCOs (including $Cu_2O$-based) is insufficient compared with well-established n-type TCOs (ITO, AZO, …), especially in terms of electrical conductivity [1,2]. This is because the valence band (VB) is derived from strongly localized oxygen 2p orbitals, leading to high effective masses of holes, $m^*$, together with high oxygen electronegativity, making it difficult to effectively increase the hole concentration by introducing shallow acceptors [2]. As the electrical conductivity, $\sigma$, is proportional to the product of charge carriers' concentration, $n$, and their mobility, $\mu$, there are generally two strategies for $\sigma$ enhancement. Firstly, one can increase $n$ by doping $Cu_2O$ by some other element. Numerous works deal with the doping of $Cu_2O$ by X (X= N, B, Ni, Li and so on). Typically, the hole concentration was increased from $\approx 10^{14}$ - $10^{16}$ cm$^{-3}$ for pure $Cu_2O$ to $\approx 10^{17}$ - $10^{19}$ cm$^{-3}$ in the case of doped $Cu_2O$:X films [3–5]. Usually, doping has a negative effect on mobility. However, the layers exhibited a higher electrical conductivity due to enhanced carrier concentration. Another possibility for improvement of electrical conductivity is to increase the hole mobility. It is often necessary to improve the hole mobility by deposition at elevated temperatures and/or by thermal post-treatment to deactivate various defects reducing the mobility. However, the classical thermal treatment (annealing) has several limitations. Firstly, the temperature used (250-1000°C) could be too high for some substrates, even for a standard soda-lime glass. Secondly, thermal annealing is mostly a long-term, non-scalable process and very often requires special conditions (vacuum or non-oxidizing atmosphere), which brings additional costs. To address this issue, a laser thermal annealing (LTA) technique is investigated as another possible route for improving the properties of various TCOs. Xu et al. used Nd:YAG laser (1064 nm, continuous) for LTA of sputtered ZnO:Al films and reported increased electrical conductivity and optical band gap after LTA [6]. The group of Yu-Chen Chen successfully used a $CO_2$ laser (power of 4.35 W) for the improvement of electrical conductivity (+19%), NIR transmittance (+16% at 1200 nm) and uniformity of ZnO:Ga films [7]. Laser irradiation of the CuO-$Cu_2O$ films with a green laser ($\lambda = 530$ nm, 10 W) for fifteen minutes was used to affect the structure of the film [8]. The laser irradiation decreased the amount of the CuO phase, and the energy band gap was shifted from 1.4 eV (as deposited) to 1.8 eV after laser treatment. In the work of Murali et al. [9], a local change of the sputtered $Cu_4O_3$ thin film to CuO was shown. Laser irradiation (532 nm) at a power density of 10 $MWcm^{-2}$ for 120 s was used.

Nevertheless, despite some works dealing with laser thermal annealing of various TCOs mentioned in the previous paragraph, only one work focused on the LTA of p-type thin-film TCO. In the work of Veron et al., $CuCrO_2$:Mg film with a delafossite structure was produced using the treatment of amorphous as-deposited films by a 405 nm solid laser [10]. However, a deeper investigation of LTA effects on properties of highly demanded p-type thin-film TCOs is still missing.

In this work, we have used a well-established high-power infrared laser for post-treatment of $Cu_2O$ thin films prepared by high-rate reactive high-power impulse magnetron sputtering (r-HiPIMS). We have systematically studied the effect of laser parameters on the optoelectrical properties of $Cu_2O$ thin films, namely electrical conductivity, the concentration of holes and their mobility, as well as on optical band gap and microstructure. We have found that LTA could be a promising way to enhance hole mobility in $Cu_2O$-based materials without the requirements of high temperature and/or a special working atmosphere.

2. Methodology

2.1 Film preparation

The Cu$_2$O films were prepared by reactive high-power impulse magnetron sputtering (r-HiPIMS) in a stainless steel vacuum chamber schematically depicted in **Fig. 1a.** The strongly unbalanced circular magnetron equipped by Cu target (area, $A_t$, of 78.54 cm$^2$, thickness of 6 mm) was fed by SPIK2000USB_S power supply (Melec GmbH). A rectangular-shaped voltage pulse at the constant length of 100 µs was applied with the repetition frequency of 100 Hz, resulting in a duty cycle, $d_c$, of 1%. The average target power, $P$, was kept constant at 500 W. The pulse-averaged target power density was then calculated as

$$S_{da} = \frac{P}{d_c \times A_t} = \frac{500}{0.01 \times 78.54} \cong 640 \; W cm^{-2} \tag{1}$$

This value is high enough to ensure a high degree of ionization of target (Cu) particles, dissociate the oxygen molecules and provide enhanced energy delivered to the growing films, which was found beneficial in our previous work related to transparent (conductive) oxides [11–13]. All depositions were carried out in Ar+O$_2$ atmosphere at constant partial pressure of argon and oxygen of 0.5 Pa and 0.3 Pa, respectively. Standard soda-lime glass (SLG, dimension of ≈ 36 × 26 mm$^2$) was used as a substrate and was cleaned in isopropyl alcohol and deionized water before deposition (10 min+10 min). The target-to-substrate distance was 100 mm, and the substrate holder was heated to 190°C. The deposition time was 70 seconds in all cases, resulting in a film thickness of 200 nm (± 5%). The corresponding deposition rate is thus ≈ 170 nm/min. After the deposition, substrates were cut into 9×9 mm$^2$ pieces suitable for Hall effect measurement using Van der Pauw geometry (see Sect. 2.4.).

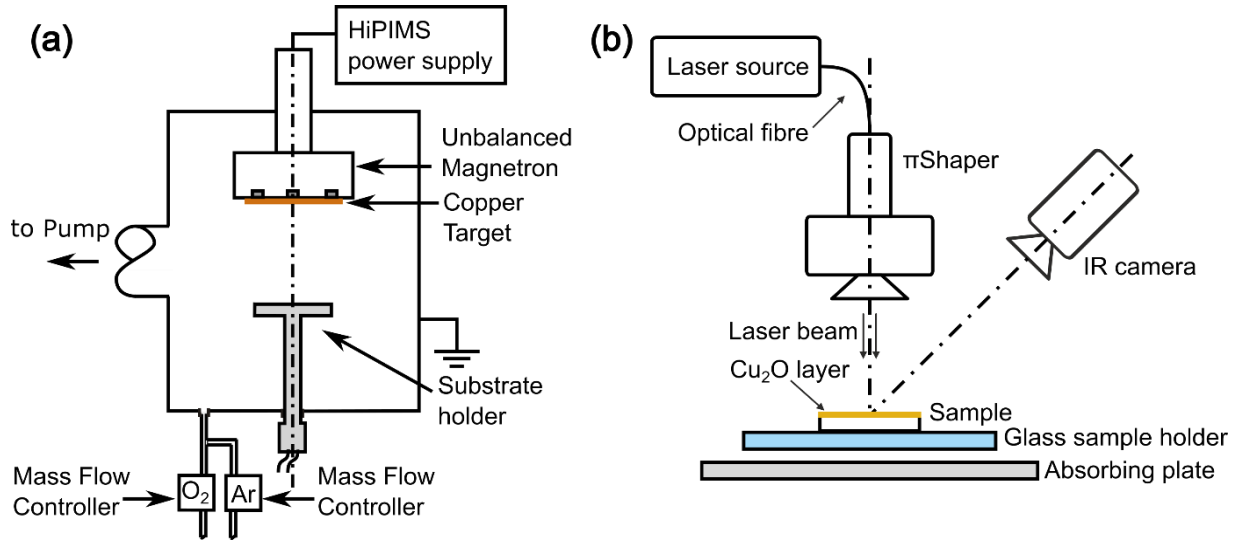

Fig. 1 (a) A schematic picture of a vacuum deposition system for preparation of $Cu_2O$ thin films, (b) a schematic picture of laser thermal annealing apparatus.

2.2 Laser thermal annealing (LTA)

For LTA, the high-power infrared laser Trumpf, TruDisk 8002(5300), having a maximum average output power of 5300 W, was used. The laser's optical output was equipped with the beam shaping system πShaper 37_34_1064 from AdlOptica. The laser device emits laser radiation at 1030 nm. The pulse length, $t_{laser}$, is adjustable from 0.3 ms to continuous operation. The pulse energy stability is ±1 %. The beam shaping system transforms the input beam with the beam parameter product 8 mm·mrad to flattop with uniform intensity distribution. This intensity distribution of laser radiation ensures a uniform irradiation of the whole sample. The system for LTA is schematically depicted in **Fig. 1b**. The Ophir Vega energy meter with the thermal sensor FL500A were used to measure the laser pulse energy and energy absorbed in the sample. The measured absorption of the laser radiation by the samples was 19 %. The power density of the laser irradiation of the sample was 3.9 kW·cm$^{-2}$ and the absorbed energies in the samples were in the range of 45 to 225 J.

2.3 Sample temperature measurement

A high-speed cooled InSb-based infrared (IR) camera FLIR A6751 with thermal sensitivity ≤ 20 mK recorded a thermal response. A lens of 25 mm, 3-5 µm, f/2.5 MWIR and a framerate of 400 Hz with a resolution of 320×256 pixels was used for the measurements. The temperature of the thin film during the LTA was evaluated from the IR camera measurement based on the measured emissivity of the sample before LTA. The maximum temperature of the sample was evaluated for each LTA parameter. The temperature time evolutions of the top of the $Cu_2O$ film and the bottom of the glass substrate collected for the shortest (75 ms) and the longest (375 ms) laser pulse are depicted in **Fig. 2.** As it is seen, the maximal temperature of the top of the $Cu_2O$ films during the LTA procedure is 170 °C and 275 °C for $t_{laser}$ = 75 ms and 375 ms, respectively. The corresponding maximal temperatures of the bottom of the SLG substrate were 115 °C and 205 °C for $t_{laser}$ = 75 ms and 375 ms, respectively. Prior to the temperature rise during the laser pulse, the constant temperature interval corresponds to the minimum of the used measuring range. For the 75 ms laser pulse the IR camera measuring range was 80 to 200 °C, and for the 375 ms 150 to 350 °C.

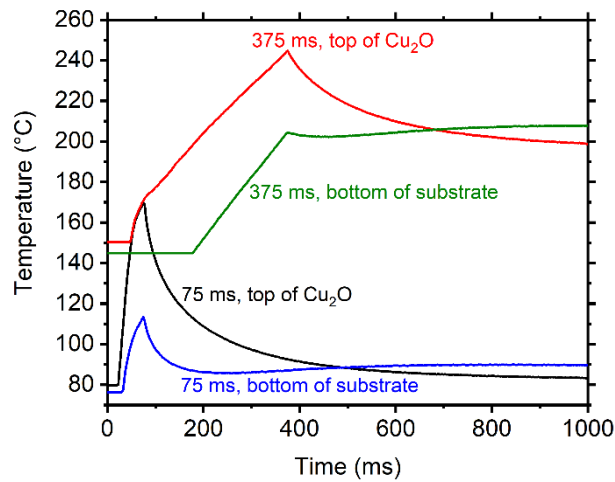

Fig. 2 The temperature time evolutions for the shortest (75 ms) and the longest (375 ms) laser pulses. The IR camera measured temperatures on the top of the $Cu_2O$ surface and the bottom of the glass substrate.

2.4 Film characterization

*Electrical properties*

The electrical conductivity of films before and after LTA was measured by a standard four-point probe method using straight-line pins configuration at room temperature. The temperature-dependent electrical resistivity, $\rho$, and Hall coefficient, $R_\text{H}$, were measured in the Van der Pauw configuration using a Variable Temperature Hall Measurement System (MMR Technologies) equipped with a Joule-Thomson refrigerator and a heating stage. Gold contacts (thickness of 50 nm) were sputter-deposited in the corners of the 9 × 9 mm² samples to improve contact between the sample and the gold spring-loaded probe tips of the Hall Measurement System. The samples were measured in the range of 200 – 380 K. The magnetic field magnitude during the Hall coefficient measurement was 1.4 T. Hole density and hole mobility were calculated, assuming the dominance of holes over electrons, as $1/(eR_\text{H})$ and $\rho/R_\text{H}$, respectively.

*Density fitting*

The temperature-dependent hole density was fitted using a model assuming two acceptor levels, as was recently done by Aggarwal et al. [14]. The acceptor levels in Cu$_2$O are attributed to simple and split Cu vacancies [15]. The charge neutrality condition implies

$$p - N_1^- - N_2^- = 0 \qquad (2)$$

where $p$ is the free hole density and $N_1^-$ and $N_2^-$ are the densities of ionized acceptor levels with activation energies $E_1$ and $E_2$ (with respect to the top of valence band), respectively. As our samples are dominantly p-type semiconductors with a band gap of around 2.4 eV, the density of free electrons can be neglected in the temperature range studied. The density of ionized acceptor can be calculated as [16].

$$N_i^- = \frac{N_i}{1+g_i\exp\left(\frac{E_v+E_i-E_F}{kT}\right)} \tag{3}$$

where $N_i$ is the density of acceptor defects, $g_i$ is the statistical weight of the acceptor level, $E_v$ is the energy of the top of valence band and $E_F$ is the Fermi energy. The density of free holes can be calculated, assuming the usual square-root energy dependence of the valence band density of states, as [16]

$$p = N_v(T)\frac{2}{\sqrt{\pi}}\int_0^\infty \frac{x^{1/2}dx}{1+e^{x-\eta}} \tag{4}$$

where $\eta = \frac{E_v-E_F}{kT}$ and $N_v(T) = 2\left(\frac{2\pi m^* kT}{h^2}\right)^{3/2}$. The effective hole mass $m^* = 0.58 m_e$ throughout this work [14,17,18] and we use the value $g = 2$ for both acceptor levels. The charge neutrality equation, including (3) and the integral formula (4) is solved numerically using the SciPy numerical algorithm package [19] so that the equilibrium value of $E_F$ is obtained for each temperature $T$. Once the Fermi level is known, the hole density $p(T)$ can be evaluated. The optimum values of $N_i$ and $E_i$ for two acceptor levels are found by fitting the model to the measured temperature-dependent hole density using the curve fitting methods from the SciPy library.

*Mobility fitting*

The temperature-dependent hole mobility is assumed to be determined by two major phenomena: grain boundary (GB) scattering, limiting the mobility at low temperatures, and trap (deep acceptor defect) scattering, limiting the mobility at high temperatures. We closely follow the work of Aggarwal et al. [14]. The mobility due to GB scattering is expressed as [20]

$$\mu_{gb} = \frac{L_{gb}e}{(2\pi m_v^* k)^{1/2}} T^{-1/2}\exp\left(-\frac{E_{gb}}{kT}\right), \tag{5}$$

where $L_{gb}$ represents the mean distance between grain boundaries (grain size) and $E_{gb}$ is the height of the energy barrier at the grain boundary. The mobility due to trap scattering is expressed as [14]

$$\mu_t = \frac{L_t e}{(3m^*k)^{1/2}} T^{-1/2} \exp(-CT), \qquad (6)$$

where $L_t$ is the mean free path between the scattering centres (accumulating both the defect density and scattering cross section) and $C$ is a parameter quantifying an exponential dependence of the product of defect density and scattering cross section on the temperature. The sample mobility can then be expressed using the Mathiessens's rule as $\mu^{-1} = \mu_{gb}^{-1} + \mu_t^{-1}$. Again, the model parameters are determined by fitting this model to the measured temperature dependence of hole mobility.

*Optical band gap*

The coating transmittance, $T$, and reflectance, $R$, at the angle of 7° from the sample normal were measured at room temperature using the Agilent CARY 7000 spectrophotometer. The value of absorption coefficient, $\alpha$, was calculated as

$$\alpha = -\frac{\ln(R+T)}{d}, \qquad (2)$$

where $d$ is the film thickness. The optical band gap was then determined from Tauc's plots, $(\alpha h\nu)^2$ vs $h\nu$, where $h$ is Planck's constant and $\nu$ the energy of incident photons.

*Microstructure*

The structure of as-deposited and laser-annealed coatings was characterized by X-ray diffraction (XRD) using a PANalytical X'Pert PRO MPD diffractometer working in the Bragg-Brentano geometry using a CuKα (40 kV, 40 mA) radiation, 0.25° divergence slit, 0.5° anti-scatter slit, 0.04 rad Soller slits, Ni filter for the CuKβ elimination and an ultrafast semiconductor detector X'Celerator. Samples were scanned over the 2θ-range from 8° to 80° with a scanning speed of 0.04 °/s. The data presented (FWHM, d-spacing) was processed by a PANalytical software package, HighScore Plus.

Top-view micrographs of the coatings were acquired in a scanning electron microscope (SEM) Hitachi SU-70 by imaging in secondary electron mode. To enhance the contrast and promote charge dissipation, the specimens were sputter-coated by 1 nm of Cr. The accelerating voltage was 5 kV.

3. Results and Discussion

3.1 Effect of LTA on microstructure

In **Fig. 3,** we can see the XRD patterns of $Cu_2O$ layers before and after the LTA procedure under different conditions. All films exhibit two prominent peaks corresponding to the c-$Cu_2O$ phase with (111) and (200) orientations. Firstly, we focus on the effect of the number of laser pulses at a constant length, $t_{laser}$ = 75 ms (**Fig. 3a) and b))**. There is not much change in the diffraction patterns as the number of pulses increases. The exception is the spectrum obtained from a sample that has been treated with sixteen laser pulses (**Fig. 3c**). Here, a significant shift of the peaks of both orientations (111) and (200) towards higher angles can already be seen. This may indicate a relaxation of the compressive stress that is very likely present after the deposition - see the detected peak positions at lower angles than the standard corresponds to. In more detail, we observe a slightly increasing FWHM with an increasing number of laser pulses (**Fig. 3a)**). This could be associated rather with an increase in lattice strain than with decreasing in grain size. Furthermore, there is a slight decrease in interplanar distances for both orientations (111) and (200) as the number of laser pulses increases.

The situation is quite different in XRD patterns in the case of laser pulse length variation (**Fig. 4c)**). With increasing pulse length, the intensity of the two prominent peaks increases and the peaks are narrower at first glance. This indicates better crystallinity of the layers (the better, the longer the laser pulse). In **Fig. 5a),** we see a significant decrease in the FWHM parameter with increasing pulse length, which can be explained by the enlargement of the crystallites. The reduction in the interplanar distance, which was also present in the series with an increasing number of pulses, is observed to be more pronounced here (**Fig. 5b**) and is probably connected with increasing lattice strain.

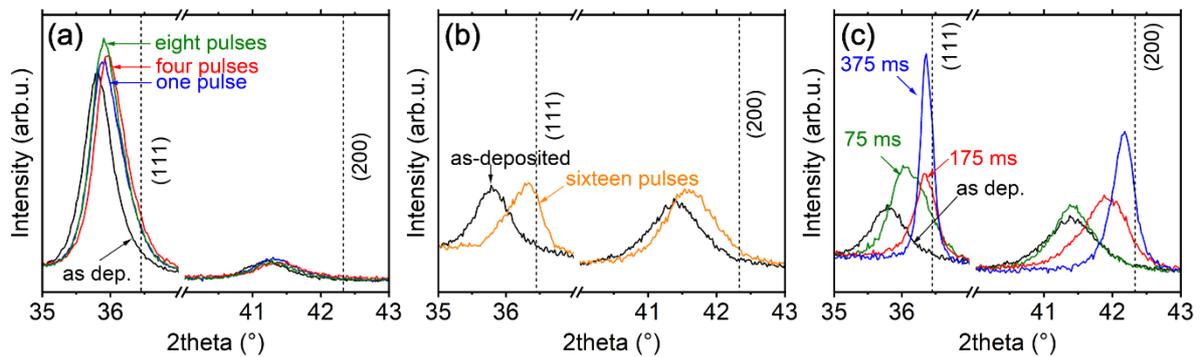

Fig. 3 (a) XRD patterns of $Cu_2O$ films before and after a different number of laser pulses ($t_{laser}$ = 75 ms); (b) XRD patterns of $Cu_2O$ films before and after sixteen laser pulses ($t_{laser}$ = 75 ms); (c) XRD patterns of $Cu_2O$ films before and after one laser pulse with different length. Dashed lines denote positions of corresponding peaks in stress-free standards.

The SEM images of the $Cu_2O$ surface before and after the LTA procedure are shown in **Fig. 6**. There is no qualitative change in surface morphology of the films treated by different numbers of laser pulses ($t_{laser}$ = 75 ms) and as deposited $Cu_2O$ film. On the other hand, there is a noticeable difference between as-deposited $Cu_2O$ morphology and film treated by laser pulse at $t_{laser}$ = 375 ms. We observe an effect of a kind of "coalescence" of grains, which is also supported by XRD patterns or by the values of FWHM parameter, as discussed above.

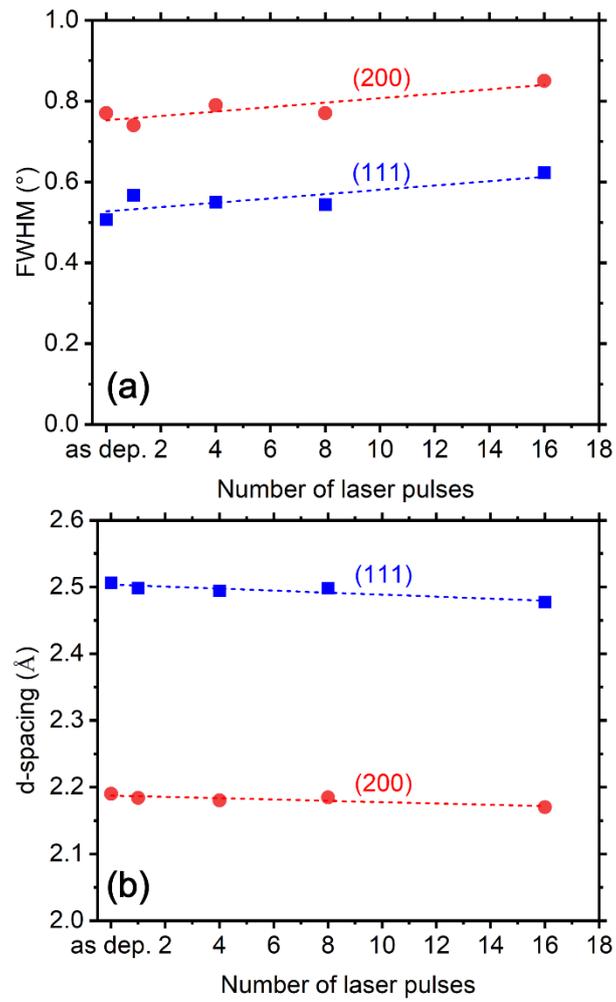

Fig. 4 (a) Full width at half maximum of corresponding diffraction peaks as a function of number of laser pulses ($t_{laser}$ = 75 ms); (b) Interplanar distance as a function of number of laser pulses ($t_{laser}$ = 75 ms). Dashed lines correspond to the linear approximations.

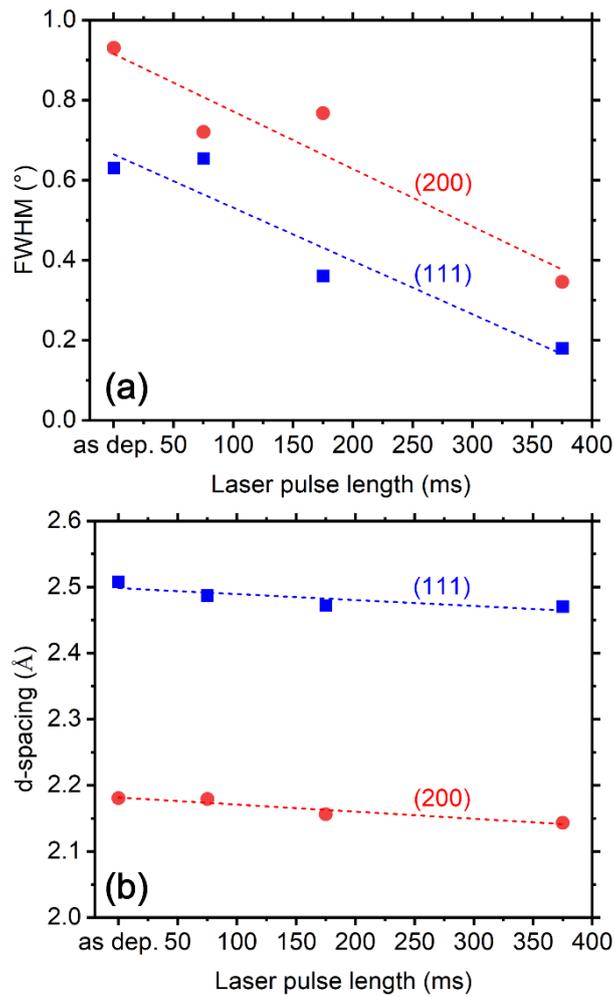

Fig. 5 (a) Full width at half maximum of corresponding diffraction peaks as a function of length of one laser pulse; (b) Interplanar distance as a function of length of one laser pulse. Dashed lines correspond to the linear approximations.

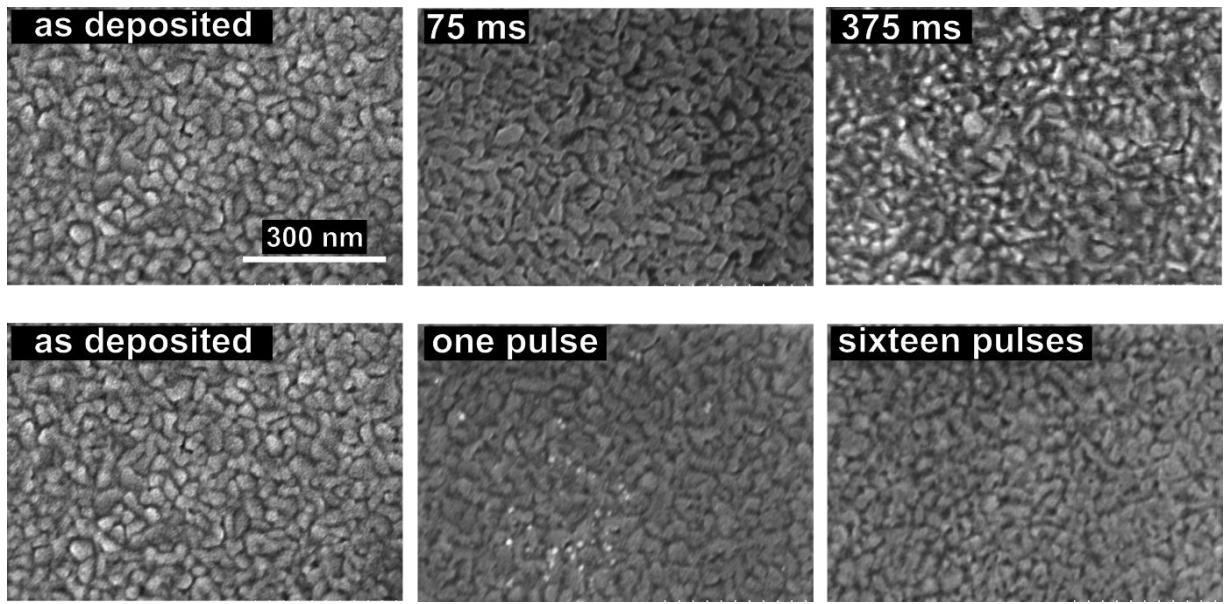

Fig. 6 Top-view SEM images of $Cu_2O$ films before and after LTA using different $t_{laser}$ (top row) and different numbers of laser pulses (bottom row)

## 3.2 Four-probe electrical conductivity measurements

To obtain an overall picture of the effect of different types of LTA, the electrical conductivity of all samples was measured at room temperature using the four-point probe method, as shown in **Fig. 7**. It is clearly seen that any LTA treatment positively affects the electrical conductivity. Interestingly, the continuously increasing number of laser pulses does not bring additional benefit in the form of continuously increasing "conductivity increase" (see **Fig. 7a)**). While for one and four pulses ($t_{laser}$ = 75 ms), the increase in electrical conductivity is 66% and 68% respectively, for sixteen pulses it is only 37%. On the other hand, we observed a monotonous increase of electrical conductivity in the case of increasing laser pulse length, where it is 51% for $t_{laser}$ = 75 ms pulse and 86% for $t_{laser}$ = 375 ms. A more detailed analysis of these phenomena is provided in the following section (Sec. 3.3).

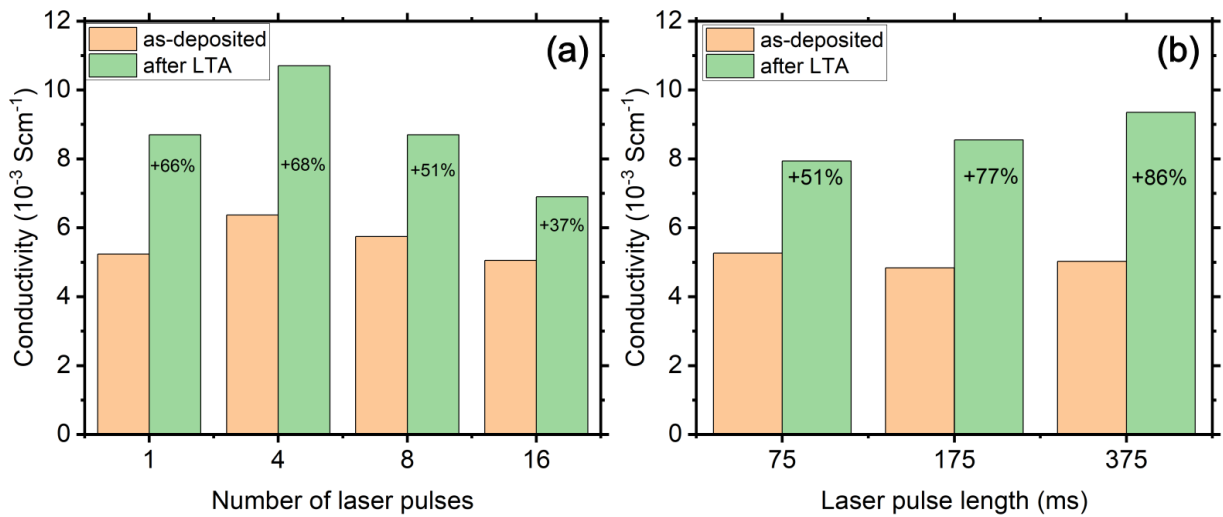

Fig. 7 Electrical conductivity of as-deposited $Cu_2O$ films and its increase after given LTA process as a function of: (a) number of laser pulses (at $t_{pulse}$ = 75 ms) and (b) laser pulse length (one pulse).

## 3.3 Hall measurements

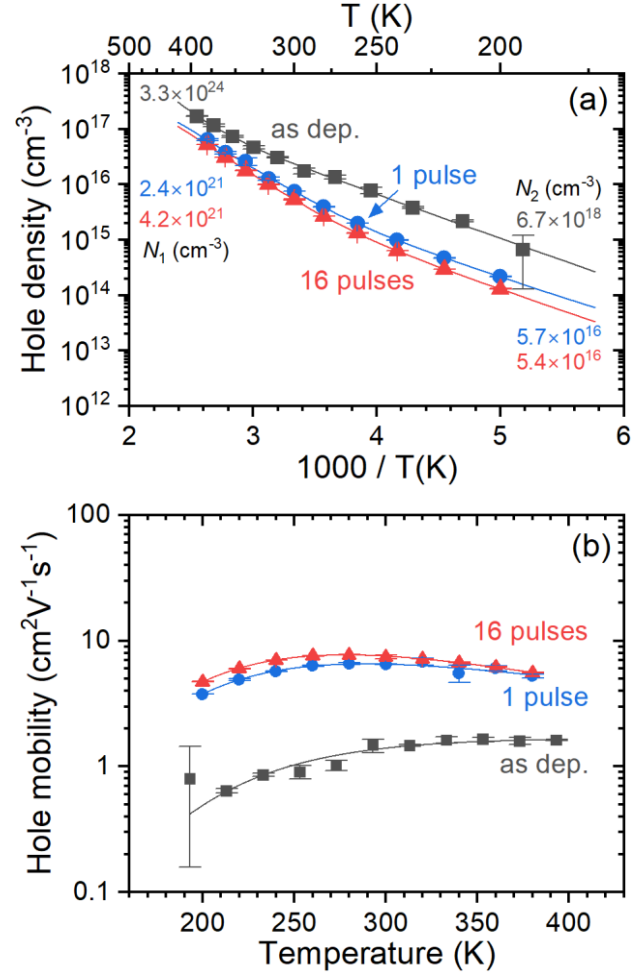

Fig. 8 The hole density (a) and the hole mobility (b) of $Cu_2O$ samples before and after the LTA procedure for various number of laser pulses ($t_{laser}$ = 75 ms). Here $N_1$ and $N_2$ are the densities of acceptor levels with activation energies $E_1$ and $E_2$ (see **Table I**)

**Fig. 8a** shows the measured hole density for various number of LTA pulses. In all cases, the hole density increases considerably with temperature. Compared to the as deposited film, after LTA, the hole density decreases by a factor of about 2 – 4. The difference in hole density between the case with 1 laser pulse and 16 laser pulses is much smaller, yet evident. The temperature dependence is not significantly altered in all three samples.

The model fits of the hole density, assuming two acceptor levels in the band gap, are shown by solid lines and the model parameters are listed in **Table I**. The densities of the defects, $N_1$ and $N_2$, associated with the two acceptor levels are also marked in the figure. Their activation energies (with respect to top of valence band) are around 0.5 eV and 0.3 eV, respectively (see **Table I** for exact values which slightly differ between samples). Thus, the defects can be characterized as deep acceptors. Because of the relatively high activation energies, the higher acceptor level is never fully ionized in the measured temperature range, and we observe the gradual increase of hole density with increasing temperature. The lower level contributes significantly to hole density only for low temperature (about for the lowest one third of the temperature range).

The fitted values of $N_1$ and $N_2$, indicate that the defect densities significantly decrease when the samples are annealed by the laser pulses. The difference between the samples after 1 and 16 laser pulses is very small with the latter sample having slightly lower $N_2$ density, corresponding to the lower hole density measured at low temperatures (200 K).

It should be noted that because the higher acceptor level is never fully ionized in the measured temperature range (no saturation region), the $N_1$ density is extrapolated from the densities in the measured temperature range. Similarly, when the lower acceptor level becomes almost fully ionized (for the higher temperatures), the higher level contributes significantly to the free hole density. Consequently, despite the ability to accurately fit the temperature dependence, we expect the error of the fitted parameters to be higher than the standard deviation given by the fitting algorithm due to imprecision of the measured data. For the defect densities, we will assume conservatively that they are accurate within one order of magnitude.

**Fig. 8b** shows the corresponding hole mobility for various number of LTA pulses. In all cases, the mobility exhibits a non-monotonous trend with a maximum that shifts depending on the LTA conditions. This is in line with the model proposed in section 2.4, i.e., there are two competing effects limiting the mobility: grain boundary scattering (low temperature) and trap center (defect) scattering (high temperature). For the as deposited sample, the mobility is the lowest and reaches a maximum of 1.6 cm$^2$/Vs at 350 K. For the samples after LTA, the mobility attains larger values in the whole temperature range with maxima of 6.6 and 7.7 cm$^2$/Vs at 280 K for one and sixteen pulses, respectively. The solid lines show that the mobility model can be well fitted to the experimental data. Although the exact values of the fitted model parameters given in **Table I** should be taken with a grain of salt, we can see that the parameters $L_{\text{gb}}$ and $L_{\text{t}}$, which correspond to effective mean free path of holes between grain boundaries and trap centres, respectively, are increasing with increasing number of LTA pulses. We can conclude that the increase in mobility can thus be mainly explained by a decrease in the defect density after LTA. It is also evident that the increase in mobility is corresponding to the decrease in hole density (Fig. 8a) by approximately the same factor (when comparing the individual samples among each other). This further strengthens the notion that the density of defects in Cu$_2$O films is the root cause for both trends.

Similarly, **Fig. 9a) and b)** show the hole density and mobility, respectively, for various lengths of LTA pulses (including the as deposited case). With the increase in laser pulse length, we observe a considerable decrease in hole density and an increase in hole mobility, again by about the same factor (when comparing the individual samples among each other). For the sample after a 375 ms laser pulse, we obtained the lowest hole density of $2 \times 10^{13}$ cm$^{-3}$ at 200 K and the highest mobility of 30 cm$^2$/Vs at 220 K. The density fitting reveals that one defect density, $N_1$, (with activation energy around 0.5 eV) does not change significantly while the other defect density, $N_2$, (with activation energy around 0.2 eV) decreases dramatically when the laser pulse length is increased from 75 to 375 ms. Again, we have to note that the exact values of fitted parameters may not be very accurate due to fitting in the limited temperature range. But the gradual decrease in hole density (especially at low temperatures) with increasing LTA pulse supports this general trend.

The fitting of hole mobility also indicates that the density of defects significantly decreases, as the effective mean free path for trap center scattering increases by almost

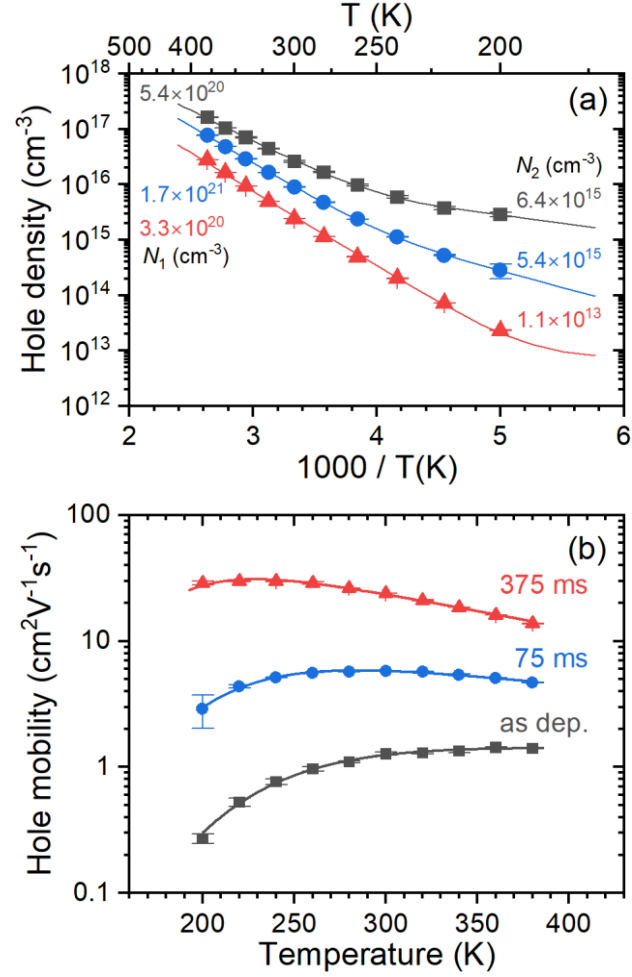

Fig. 9 The hole density (a) and the hole mobility (b) for Cu$_2$O samples before and after the LTA procedure under various laser pulse length. Here $N_1$ and $N_2$ are the densities of acceptor levels with activation energies $E_1$ and $E_2$ (see **Table I**)

two orders of magnitude when the LTA pulse length is increased. In contrast to the effect of laser count, the mean free path for grain boundary scattering also increases by almost two orders of magnitude, reflecting the dramatic increase in mobility at low temperatures. This qualitatively agrees with increasing of grains sizes with increasing $t_{\text{laser}}$ as discussed in Sect. 3.1. Thus, we can conclude that increasing the length of LTA pulses leads to a decrease in defects in Cu$_2$O films and increase in grain sizes which lead to enhancement of hole mobility and decrease in hole density.

Finally, we can conclude that the increase in mobility is more pronounced for an increased pulse length than for an increased number of pulses (see almost identical values for one and sixteen pulses in Fig. 8). Increasing the laser pulse length is thus more effective in increasing hole mobility than the number of laser pulses.

Table I Model parameters for fitting the temperature dependence of hole mobility and hole density for samples subject to various number of laser pulses and length of laser pulses. Here, $N_1$, $N_2$ and $E_1$, $E_2$ are densities and activation energies of acceptor levels, respectively, $L_{\text{gb}}$ is the mean grain size, $E_{\text{gb}}$ is the grain boundary energy barrier height, $L_{\text{t}}$ is the mean free path for trap scattering and $C$ is a parameter for its exponential temperature dependence,

| Pulse count | $N_1$ | $E_1$ | $N_2$ | $E_2$ | $L_{gb}$ | $E_{gb}$ | $L_t$ | $C$ |
|---|---|---|---|---|---|---|---|---|
| - | cm$^{-3}$ | eV | cm$^{-3}$ | eV | nm | eV | nm | $10^{-3}$ K$^{-1}$ |
| as dep. | $3.3 \times 10^{24}$ | 0.71 | $6.7 \times 10^{18}$ | 0.29 | 12 | 0.10 | 0.12 | 0 |
| 1 | $2.4 \times 10^{21}$ | 0.51 | $5.7 \times 10^{16}$ | 0.26 | 40 | 0.09 | 1.4 | 3.7 |
| 16 | $4.2 \times 10^{21}$ | 0.54 | $5.4 \times 10^{16}$ | 0.28 | 45 | 0.08 | 2.6 | 5.1 |
| **Pulse length** | $N_1$ | $E_1$ | $N_2$ | $E_2$ | $L_{gb}$ | $E_{gb}$ | $L_t$ | $C$ |
| ms | cm$^{-3}$ | eV | cm$^{-3}$ | eV | nm | eV | nm | $10^{-3}$ K$^{-1}$ |
| as dep. | $5.4 \times 10^{20}$ | 0.40 | $6.4 \times 10^{15}$ | 0.12 | 50 | 0.14 | 0.09 | 0 |
| 75 | $1.7 \times 10^{21}$ | 0.48 | $5.4 \times 10^{15}$ | 0.21 | 300 | 0.12 | 0.7 | 2.4 |
| 375 | $3.3 \times 10^{20}$ | 0.50 | $1.1 \times 10^{13}$ | 0.17 | 2000 | 0.11 | 6 | 5.2 |

Table II Comparison of experimental conditions and properties of pure $Cu_2O$ films prepared by different methods onto amorphous substrates. Here, $T_{dep}$ and $T_{post}$ are the temperatures of deposition and post-treatment, respectively. $a_D$ is deposition rate, $\tau_{post}$ is post-treatment duration. $\sigma$ and $\mu$ are electrical conductivity and hole mobility at room temperature, respectively. SLG and BSG mean soda-lime and borosilicate glass, respectively.

| Method | Substrate | $T_{dep}$ (°C) | $a_D$ (nm/s) | $T_{post}$ (°C) | $\tau_{post}$ (s) | $\sigma$ (S/cm) | $\mu$ (cm$^2$/Vs) |
|---|---|---|---|---|---|---|---|
| **HiPIMS (this work)** | **SLG** | **190** | **2.85** | **260** | **≈0.4** | **≈0.01** | **24** |
| HiPIMS [21] | BSG | ambient | 0.7 | - | - | ≈0.02 | 32 |
| HiTUS [22] | Corning g. | 100 | 0.6 | 250 | 3600 | ≈0.01 | 10 |
| DCMS [17] | SLG | 327 | 0.07 | - | - | ≈0.01 | 43 |
| DCMS [23] | SLG | 600 | 0.5 | - | - | ≈0.01 | 54 |
| RFMS [24] | Quartz | 400 | 0.5 | 1000 | 180 | ≈0.007 | 58 |

**Table II** gives an overview of the electrical properties of $Cu_2O$ thin films deposited under different conditions on amorphous substrates. We dare to say that the results published in this work, i.e. the combination of r-HiPIMS and the subsequent treatment of $Cu_2O$ by infrared laser, are very competitive. Especially if we consider the very high deposition rate we achieved (2.85 nm/s) combined with a relatively high mobility after LTA (up to 24 cm$^2$/Vs). At the same time, the substrate and layer temperature remain relatively low (maximum 260 °C for less than 375 ms) compared to conventional annealing. Moreover, it is a scalable and selective method that did not require any special working atmosphere in this case.

3.4 Optical band gap

The Tauc's plots for $Cu_2O$ films before and after LTA procedure under various number of laser pulses and the values of the optical band gap, $E_g$, derived from them are shown in **Fig. 10**. The values of $E_g$ increases from 2.38 eV (as deposited) up to 2.47 eV for the $Cu_2O$ film treated by four laser pulses, however, then decreases to 2.44 eV and 2.40 eV for eight and sixteen laser pulses, respectively. On the other hand, in the case of $Cu_2O$ films treated by one laser pulse with different $t_{laser}$, we observed an increase of $E_g$ with increasing $t_{laser}$ up to 2.48 eV. Many phenomena affect the optical band gap of $Cu_2O$. Koshy et al. observed a rise in $E_g$ of $Cu_2O$ film after annealing at 100°C in the ambient atmosphere due to a decrease in defect density caused by the presence of oxygen in the annealing atmosphere and due to Burstein-Moss shift (shifting of the unoccupied states towards higher energies in the allowed bands) [25]. Kumar also reported a blue shift of absorption edge and a corresponding increase of optical band gap after annealing of $Cu_2O$ at 900°C and attributes this to, among other things, better crystallinity of the layers [26]. In the theoretical paper published by Visibile et al., the influence of strain on the value of the $Cu_2O$ optical band gap is discussed using ab-initio calculations [27]. Authors reported that both tensile and compressive strain can result in a decrease in the band gap. In the context of above information, we suggest: Just after one short (75 ms) laser pulse, there is a significant decrease of hole concentration accompanied with increase in hole mobility (see **Fig. 8**) indicated the decreasing defect density. This is the reason for increasing $E_g$ up to 2.47 eV for the film treated by four pulses. However, further increasing of laser pulses number at the same $t_{laser} = 75$ ms does not lead to a significant further decrease of defect concentration (see very close values of hole mobility and concentration for one and sixteen pulses, **Fig. 8**). But on the other hand, there is a continuous increase in the FWHM parameter with increasing of the number of laser pulses. There is no reason for decreasing the size of crystal grains after the LTA procedure with an increasing number of laser pulses, as one could conclude from the increasing FWHM parameter (**Fig. 4a**). Thus, we suggest the increasing FWHM is rather connected with increasing lattice strain. So, for a higher number of short pulses, the lattice strain becomes the dominant effect, resulting in a slight decrease of the optical band gap very close to the as deposited value. The increasing laser pulse length leads to a significant decrease in FWHM (see **Fig. 5a)**), and although without a deeper analysis the influence of grain size and lattice strain cannot be distinguished from the FWHM parameter, we suggest the grains become larger for annealing with longer laser pulse length. This is supported by the huge difference in room-temperature hole mobility of 5.8 and 23.8 cm$^2$/Vs for $t_{laser}$ = 75 ms and 375 ms, respectively which cannot be explained without a significant decrease of grain boundaries volume (see Sec.3.3) and also by surface morphology (**Fig. 6**). Thus, the enhanced crystallinity together with decreasing defect concentration are responsible for $E_g$ increase in this case.

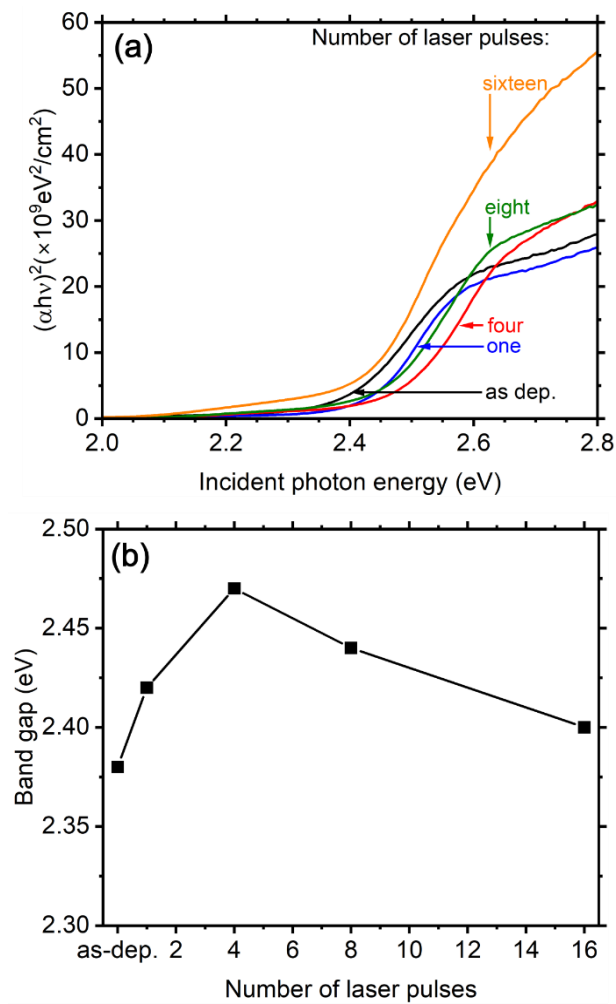

Fig. 10 The Tauc's plots (a) and the extracted values of optical bandgap of Cu$_2$O films before and after LTA under different laser pulses (b).

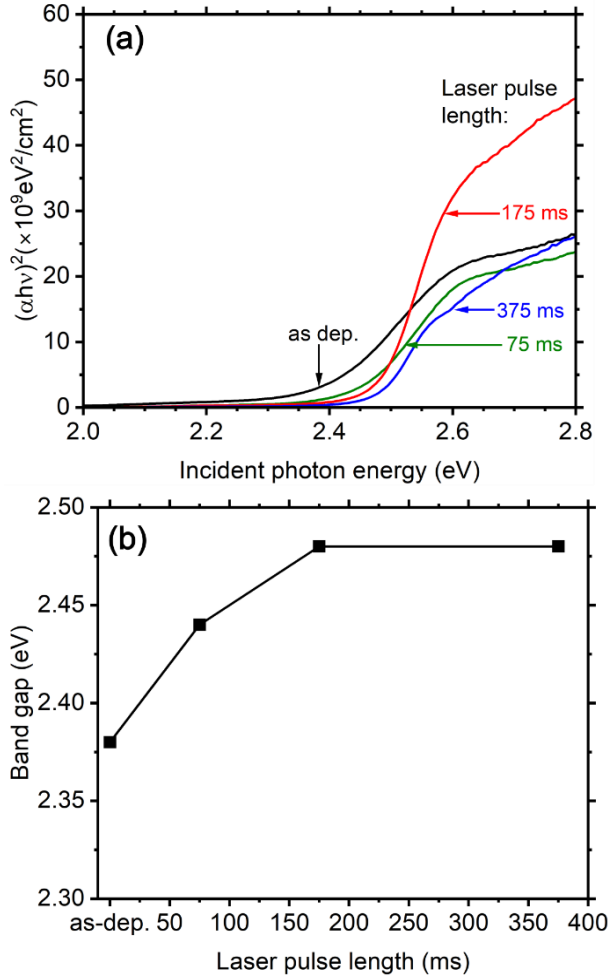

Fig. 11 The Tauc plots (a), and the extracted values of optical bandgap of $Cu_2O$ films before and after LTA procedure under different $t_{laser}$ (b).

4. Conclusions

The high-rate (≈ 170 nm/min) reactively sputtered $Cu_2O$ films was successfully post-treated by high-power infrared laser thermal annealing (LTA). We showed a positive effect of LTA on optical and electrical properties. The $Cu_2O$ film post-treated by one 375 ms pulse exhibit significantly better crystallinity and an 86% increase in electrical conductivity compared with the as-deposited sample. Moreover, there is a significant increase in room-temperature hole mobility compared with the as-deposited state ( ≈ 1 vs. 24 $cm^2$/Vs) and optical band gap (2.38 vs 2.48 eV). We suggest that the main reasons are the decrease in defect density and increase in crystallinity after the LTA procedure similarly as it is observed in the case of a conventional thermal annealing. However, using LTA, it was achieved at a lower temperature and several orders of magnitude lower treatment time. Overall, we found that laser thermal annealing through a well-established and easily up-scaled infrared laser could be a promising way for hole mobility enhancement of $Cu_2O$ based materials.

**CRediT authorship contribution statement**

**Jiří Rezek**: Conceptualization, Methodology, Experiment, Investigation, Visualization, Writing – original draft, Writing – review & editing. **Martin Kučera:** Investigation, Experiment **Tomáš Kozák:** Methodology, Investigation, Visualization, Writing – original draft, Writing – review & editing. **Radomír Čerstvý:** Investigation. **Aleš Franc**: Experiment. **Pavel Baroch:** Supervising, Conceptualization.

**Acknowledgement**

This work was supported by the project Quantum materials for applications in sustainable technologies (QM4ST), funded as project No. CZ.02.01.01/00/22_008/0004572 by Programme Johannes Amos Commenius, call Excellent Research. We thank Dr. Stanislav Haviar for taking SEM images of the layers.